\documentclass[a4paper,1pt,twocolumn]{article}
\usepackage[dvips]{graphicx}
\usepackage{amssymb}
\usepackage{amsmath}
\usepackage{setspace}
\usepackage{setspace}
\usepackage[super]{cite}
\input{tcilatex}

\begin{document}
 \begin{spacing}{1.}

\noindent {\Large\textbf{The nanofluidics can explain   ascent of water  in tallest trees}}
\\

\noindent \textbf{{Henri Gouin}\footnote{henri.gouin@univ-amu.fr\\
M2P2,\, C.N.R.S.\, U.M.R.  7340 \&  Universit\'e d'Aix-Marseille,\\  Avenue Escadrille Normandie-Niemen,  13397 Marseille Cedex 20 France}}
\\

\noindent \textbf{{Abstract.}}\\

\noindent \textbf{\footnotesize  In \emph{Amazing numbers in biology}, Flindt reports  a giant, 128 meter-tall  eucalyptus,
and
   a 135 meter-tall sequoia  \cite{Flindt}\,. However, the
explanation of the maximum altitude of the crude sap ascent and consequently  the main reason of the maximum size that trees can reach is   not well
 understood  \cite{Koch}\,.\\ According to tree species, the crude sap is driven in xylem microtubes with diameters ranging between 50 and  400  micrometers. The sap  contains diluted salts
  but  its physical properties are roughly those of  water; consequently,   hydrodynamic,  capillarity  and osmotic pressure
  yield a crude sap ascent of  a  few tens of meters  only \cite{Zimm}\,.\\
 Today, we can propound a new   understanding of the
ascent of sap to the top of very tall trees thanks to a new comparison between experiments associated with the
\emph{cohesion-tension theory} \cite{Dixon} and the \emph{disjoining pressure  concept} \cite{Derjaguin}\,.\\
  Here we show that   the pressure in   the water-storing tracheids of leaves can be strongly negative whereas the pressure in the xylem microtubes of   stems may remain positive  when,
 at high level, inhomogeneous liquid nanolayers  wet the xylem walls of microtubes.\\  The nanofluidic model of crude sap  in tall trees \cite{gouin4} discloses a stable sap layer  up to an altitude where the \emph{pancake layer} thickness \cite{deGennes1} coexists with the dry xylem wall and corresponds to the maximum size of tallest trees.
       In  very thin layers,  sap  flows are widely more significant than those obtained with classical Navier-Stokes models and consequently are able to refill stomatic cells when phloem embolisms supervene.
\newline
   These results  drop an inkling that the disjoining pressure is an efficient tool  to study biological liquids in contact with substrates at a nanoscale range.}

\noindent\textbf{---------------------------------------------------------}\\

\begin{figure}[h]
\begin{center}
\includegraphics[width=8cm]{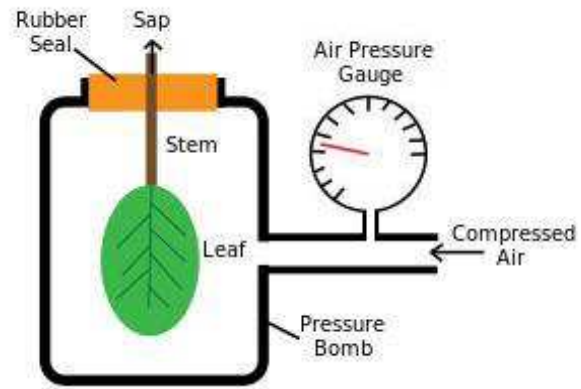}
\end{center}
  \begin{spacing}{0.785}\caption{ \footnotesize \textbf{Sketch of the Scholander pressure chamber} (or Scholander pressure bomb) \cite{Scholander} .  A leaf attached to a stem is placed inside a sealed chamber. Compressed air is  slowly added to the chamber. As the pressure increases to a convenient level, the sap is forced out of the xylem and is visible at the cut end of the stem. The   required pressure is  opposite and of equal magnitude to the water pressure in the water-storing  tracheids in the leaf.} \label{fig1}
  \end{spacing}
\end{figure}

\small
The most classical explanation of the sap ascent phenomenon in tall trees  is given by  the cohesion-tension theory   propounded in
1894 by Dixon and Joly  \cite{Dixon}\,, followed by a
quantitative analysis of the sap motion proposed by van der Honert in 1948 \cite{vanderHonert}\,:
  According  to this theory, the crude sap  fills  water-tight microtubes of dead xylem cells and its transport is due to a gradient of negative pressure  producing the traction necessary to lift water against gravity. A main experimental checking  on the cohesion-tension theory comes from an apparatus called   \emph{Scholander pressure chamber}  (see Fig.\,\ref{fig1}).
The decrease in the negative pressure is related to the closing of the aperture of microscopic stomata in leaves through which water vapour is lost by transpiration.

 Nonetheless, several objections question  the cohesion-tension theory:\\
We first refer  to
the well-known book by M.H. Zimmermann \cite{Zimm}\,. He said:
 "The heartwood is referred to as a wet wood. It may
contain liquid under positive pressure while in the sapwood the
transpiration stream moves along a gradient of negative pressures.
Why is the water of the central wet core not drawn into the sapwood?
Free water, i.e. water in tracheids, decreases in successively older
layers of wood as the number of embolized tracheids increases. The
heartwood is relatively dry i.e. most tracheids are embolized.
 It is rather ironic that a wound
in the wet wood area, which bleeds liquid for a long period of time,
thus appears to have the transpiration stream as a source of water,
in spite of the fact that the pressure of the transpiration stream
is negative most of the time!
It should be quite clear by now that a drop in xylem pressure below
a critical level causes cavitations and normally puts the xylem out
of function permanently. The cause of such a pressure drop can be
either a failing of water to the xylem by the roots, or excessive
demand by transpiration."\\
At great elevation in  trees, the value of the negative pressure increases the risk of cavitation and
 consequently, the formation of embolisms may cause a breakdown of the continuous column of sap inducing the leaf death. The
crude sap is a liquid bulk with a superficial tension $\sigma$ lower
than the superficial tension of pure water which is about $72.5$ cgs at
20${{}^\circ}$ Celsius; if we consider a microscopic gas-vapour
bubble  with a diameter $2 R$ smaller than
xylem microtube diameters, the difference between the gas-vapour pressure
and the liquid sap pressure is expressed by the Laplace formula
$\displaystyle P_v-P_l = 2\,\sigma/R$; the vapour-gas pressure is
positive and
 consequently unstable
bubbles must appear  when $\displaystyle
  R\geq
- 2\,\sigma/P_l$. For a negative pressure  $P_l =-0.6$ MPa in the
sap, we obtain  $R \geq 0.24\, \mu m$;  then, when all the vessels are tight filled, germs
naturally pre-existing in   crude water
 may spontaneously  embolize the tracheids.\\
 Haberlandt \cite{Haberlandt} described water-storing tracheids in leaves; they
are roundish in shape, and located either at the tips of the veins or
detached from transporting xylem. In more recent papers they
have been called  \emph{tracheid idioblats} \cite{Foster}\,. The
spacing considered in Pridgeon's paper \cite{Pridgeon} is about $2\, \mu m$ or less at the top of trees as suggested on Fig.\,2 of the paper by Koch \emph{et al} \cite{Koch} and has a good size to
prevent  cavitation for nucleus germs of the same
order of magnitude.\\
  No vessels are
continuous from root to stem, from stem to shoot, and from shoot
to petiole. The vessels do not all run neatly parallel and form a
network  generally  up a few centimeters long. The
ends usually taper out; it is very important for the understanding
of water conduction to realize that the water does not leave a
vessel in axial direction through the very end but laterally along a
relatively long stretch where the two vessels, the ending and the
continuing  ones, run side by side.\\
The vascular bundles of some leaves are surrounded
by a bundle sheath, containing a suberized layer comparable to the one
of the Casperian strip in the roots \cite{0'Brien}\,. This seal
separates the apoplast into two compartments, one inside and the
other one outside the bundle sheath. The two areas are only connected by the
plasmodesmata that connect living cells.
The pressure in the intact, water-containing neighbouring tracheids,
may still be negative; a considerable pressure drop therefore exists
across the pit membranes. Pressure chamber measurements cannot be
considered as pressure values of the stem xylem without special
precautions, simply because they are taken elsewhere.
Hydraulically then, the leaf is very sharply separated from the
stem. The wet wood area of elms appears to act like  a single, giant
osmotic cell that is separated   from
the sapwood area by a semi permeable  membrane. This can be visualized somewhat like a Traube
membrane, as early plant physiologists called it \cite{Traube}\,.\\
 An other   objection
in the perfect confidence  to the
cohesion-tension theory was the  experiments by Preston
\cite{Preston} who demonstrated that tall trees survived by
overlapping double saw-cuts made through the cross-sectional area of
the trunk to sever all xylem elements. This result,  confirmed   by
several authors (e.g.
 \cite{Mackay}), does not seem  in agreement with
the possibility of strong negative pressures in the water-tight microtubes.
Using a xylem pressure probe, Balling \emph{et al} \cite{Balling} showed that, in many
circumstances, this apparatus does not measure any water tension
 \cite{Tyreeetal}\,.

As a consequence of these various pieces of evidence,  the main question is:\\
The negative pressure measured by the Sholander pressure chamber being the pressure in  the water-storing tracheids, is it possible that the pressure in the xylem microtubes of the stem remains positive?

\begin{figure}
\begin{center}
\includegraphics[width=8cm]{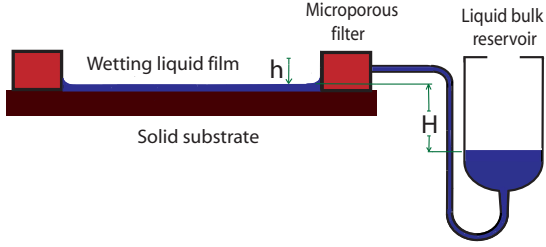}
\end{center}
 \begin{spacing}{0.7} \caption{{\footnotesize \textbf{The Sheludko apparatus .} Schematic diagram of the technique to determine the disjoining
pressure isotherms of wetting films on a solid substrate: a circular
wetting film is formed on a flat substrate to which a microporous
filter is clamped. A pipe connects the filter filled with the liquid
to a reservoir containing the    liquid bulk  that can be moved
by a micrometric device.   The disjoining pressure is
equal to $\Pi = (\rho_{l_b}-\rho_{v_b})\,g\,H$, where $g$ is the
acceleration of gravity and  $\rho_{l_b},\, \rho_{v_b}$ are the densities in the liquid and the vapour-gas bulks, respectively (From\cite {Derjaguin}  page
332)}.}\label{fig2}
 \end{spacing}
\end{figure}

A positive answer to this question comes from the concept  of disjoining pressure   which is able to interpret very thin vertical films of liquid wetting solid substrates.
The
simplest
  review  relative to the disjoining pressure
is presented in the well-known monograph  by   Derjaguin \emph{et al} \cite{Derjaguin}\,.
An apparatus measuring the disjoining pressure denoted $\Pi$ is due to Sheludko \cite{Sheludko}\,.  The apparatus is schematically described on Fig.\,\ref{fig2} and allows   to understand the disjoining pressure concept.
Fluids and solids are at the same  temperature. The film is thin enough for the
gravity effect across  the layer to be neglected. The hydrostatic pressure in a thin
liquid layer included between a solid substrate and a vapour-gas bulk differs
from the pressure in the  liquid bulk contained in the reservoir. The forces arising from the thinning of a film of uniform
thickness   $h$
yield the disjoining pressure as a function of $h$ ($\Pi = \Pi(h)$).
 The disjoining pressure is equal to the difference between
  the pressure $P_{{v_b}}$ through the interfacial surface of the thin liquid layer
  (which is the total pressure of gases and   \emph{vapour  bulk}) and the
  pressure $P_{l_b}$ at the top of the liquid   bulk  with dissolved gases contained in the reservoir
  from which the thin liquid layer extends:
\begin{equation*}
\Pi  = {P}_{{v_b}}-{P}_{l_b}.  \label{disjoiningpressure}
\end{equation*}
As a
vapour-gas pressure, the pressure $P_{{v_b}}$  is always positive; depending on H-values, the pressure $P_{l_b}$
  may be negative.
\newline

These comments illuminate an original comparison between the Sholander pressure chamber experiment presented in Fig.\,\ref{fig1} and the apparatus proposed by Sheludko in Fig.\,\ref{fig2} and make a total change in the interpretation of Sholander pressure chamber data for the tallest trees:\newline
In the Sholander pressure chamber, the cohesion-tension theory assumes that, at a given level, liquids entirely convey the pressure as  is the case for incompressible fluids. This assumption is for thin layers at odds with the Sheludko experiment  where the disjoining pressure highlights a strong difference between liquid bulk and thin layer pressures.
The xylem of stem walls is associated with the solid substrate and the water-storing tracheids in leaves are associated with the liquid bulk reservoir of Sheludko's experiment.\newline In our new interpretation, the negative pressure measured in the Sholander pressure chamber is the pressure in water-storing tracheids corresponding to the liquid bulk reservoir of Sheludko  while the pressure in the xylem microtubes of stems remains positive as in the wetting thin layer of Sheludko's experiment when H-value is equal to the level in the tree.
\newline
\begin{figure}
\begin{center}
\includegraphics[width=7.5cm]{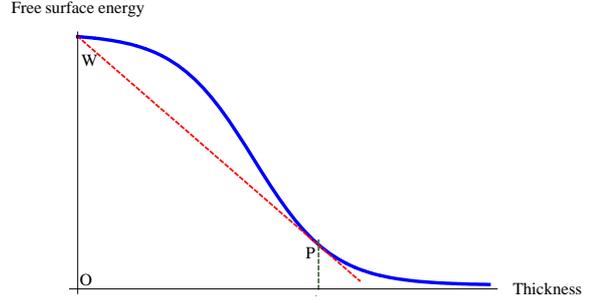}
\end{center}
\begin{spacing}{0.7 }
\caption{{\footnotesize \textbf{The pancake layer thickness .} The construction of the tangent to  curve $G(h)$ issued from point  W  of coordinates  $(0,G(0))$  yields point  P; point
W  is associated with a high-energy surface  of the dry wall and point  P  is associated with   the pancake
thickness  h$_{\rm p}$\cite{deGennes1} .}}\label{fig3}
\end{spacing}
\end{figure}
As proven  in \cite{Derjaguin}\,, Chapter 2,
 the   Gibbs free energy per unit area  of
the  liquid layer denoted $G$ can be expressed as a function of
 $h$ thanks to the relation:
\begin{equation*}
{G (h)} = \int_{h}^{+\infty} \Pi(u)\,du,\label{Gibbs2}
\end{equation*}
where $h = 0$  is associated with a dry wall in contact with the vapour-gas
bulk and  $h = +\infty$ is associated with a wall in
 contact with  the  liquid bulk  when the value of  $G$  is equal to $0$.
 The coexistence of two film segments with different thicknesses is a phenomenon which can be represented by
 the equality of chemical potential and superficial tension of the two films \cite{Derjaguin}\,.
A spectacular case
  corresponds to the coexistence of a liquid film of thickness $h_{\rm p}$ and the dry solid wall;    the liquid film is the  so-called  pancake  layer and corresponds  to the minimal thickness for
which a stable wetting film damps a solid wall   \cite{Derjaguin,deGennes1}\,. This minimal thickness  $h_{\rm p}$ verifies:
\begin{equation}
{G}(0) =  {G}(h_{\rm p})+ h_{\rm p} \Pi(h_{\rm p}). \label{pancake thickness}
\end{equation}
The geometrical
interpretation of Eq. (\ref{pancake thickness}) is proposed on Fig.\,\ref{fig3}.
 \newline
Now, we consider a film of thickness $h_x$ at level $x$.   Only liquid water films of thicknesses $h_x >$ ${h_{\rm p}}$ are stable.
The disjoining pressure of the mixture of water and perfect gas is the same
as for a single  liquid far under its critical point \cite{Derjaguin,gouin4}\,. Calculations limited to overstrained liquid layers of a few nanometers thick yield
 the disjoining
pressure in the approximated form \cite{gouin3,gouin5}\,:
\begin{eqnarray}
\Pi (h_x)\simeq \quad\qquad\qquad\qquad\qquad\qquad\qquad\qquad\qquad \notag \\
\frac{2\,c_{l}^{2}}{\rho _{l}}\left[ (\gamma _{1}-\gamma
_{2}\rho _{l})(\gamma _{2}+\gamma _{3})e^{h_x\tau }+(\gamma _{2}-\gamma
_{3})\gamma _{2}\rho _{l}\right]  \notag \\
 \times \frac{\left[ (\gamma _{2}+\gamma _{3})\gamma _{2}\rho _{l}+(\gamma
_{1}-\gamma _{2}\rho _{l})(\gamma _{2}-\gamma _{3})e^{-h_x\tau }\right] }{%
\left[ (\gamma _{2}+\gamma _{3})^2 e^{h_x\tau
}-(\gamma_{2}-\gamma _{3})^2 e^{-h_x\tau }\right] ^{2}}
,  \label{DerjaguineDP}
\end{eqnarray}
where     $\rho_l$ is the liquid density in normal conditions,  $c_l\,$ is the isothermal sound velocity in the liquid water bulk, $\gamma _{1}$, $\gamma _{2}$, $\gamma _{3}$ and $\tau$ are positive constants given
by the mean field molecular theory \cite{rowlinson,gouin1}\,:
 \begin{eqnarray*}
\gamma _{1}=\frac{\pi c_{ls}}{12\delta ^{2}m_{l}m_{s}}\;\rho _{sol},\quad
\gamma _{2}=\frac{\pi c_{ll}}{12\delta^2 m_{l}^{2}},\\ \gamma _{3} =
c_l\sqrt{\frac{2\pi c_{ll}}{3\sigma_l \,m_{l}^{2}}}, \quad \tau =   c_l\sqrt{\frac{3\sigma_l \,m_{l}^{2}}{2\pi c_{ll}}}.
\end{eqnarray*} Constants $c_{ll}$ and $c_{ls}$ are  associated with
Hamaker coefficients of interaction of liquid \emph{vs} liquid and liquid \emph{vs} solid \cite{Israel}\,; $\sigma_l$   denotes the fluid
 molecular diameter and  $\delta= \frac{1}{2}(\sigma_l+\sigma_s)$, where $\sigma_s$ denotes the solid
 molecular diameter; $%
m_{l}$, $m_{s}$ denote the masses of fluid and solid molecules; $%
\rho _{sol}$ denotes the solid density.  Expression (\ref{DerjaguineDP})   differs from Lifshitz's relation  \cite{Lifshitz} where the disjoining pressure of microscopic layers of liquid, assumed to be incompressible, has a behavior in the form $%
\Pi(h) \sim h^{-3}$.
\\ To obtain the pancake thickness corresponding to the smallest
thickness of the liquid layer, we draw the graphs of $G(h_x)$  and $\Pi(h_x)$ when $h_x \in
[({1}/{2})\,\sigma_l,\ell]$, where $\ell$ is a length of a few tens of
 \aa ngstroms.
Let us note that $d = 1/\tau$ is the natural
 reference length scale.    For a few nanometers,
the film thickness is not exactly $h_x$; we must add the thickness estimated   at $2\,\sigma_l$ of the liquid
part of the liquid-vapour interface bordering the liquid layer  and the
nanolayer thickness is approximatively $e_x \approx h_x+ 2\, \sigma_l$ \cite{rowlinson}\,.

Our aim is now to point out a numerical example such that previous results provide a
value of maximum height for a vertical water film wetting a  wall of
xylem.
 \begin{figure}[h]
\begin{center}
\includegraphics[width=7.9cm]{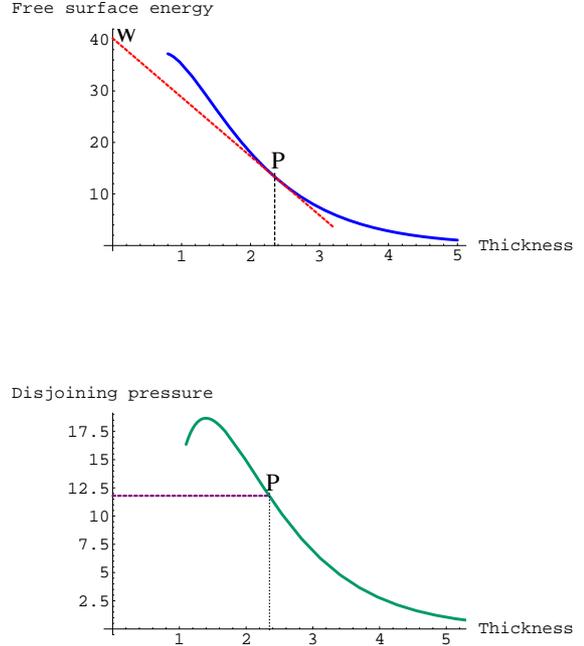}
\end{center}
\caption{\footnotesize \textbf{The maximum height of trees .} {\emph{Upper graph}}:  $G(h_x)$-graph. The unit of $x-$axis graduated by $h_x$  is $%
d=2.31\times 10^{-8}$ \texttt{cm} ; the unit of $y-$axis is one
c.g.s. unit  of surface tension. {{\emph{Lower graph}}: $\Pi(h_x)$-graph. The unit of $x-$axis graduated by $h_x$
is $d=2.31\times 10^{-8}$ \texttt{cm}; the unit of $y-$axis is one
atmosphere. }}
\label{fig4}
\end{figure}
 We considered water at $20 {%
{}^\circ}$ Celsius. The experimental estimates of
coefficients are expressed in  {c.g.s. units} \cite{Israel,Handbook}\,:
$\rho_l =0.998$, $c_l = 1.478\times 10^{5}$, $c_{ll}=1.4\times 10^{-58}$, $%
\sigma _{l}=2.8\times 10^{-8},$ $%
m_{l}=2.99\times 10^{-23}$.  We deduce
 $\gamma _2= 54.2$, $\gamma _3 = 506$,  $d = 2.31 \times 10^{-8}$. \newline
We consider the Young contact angle  between the xylem wall and the liquid-vapour water interface as $\theta \approx 50{{}^\circ}$   (this value is an arithmetic average  of different Young angles proposed in the literature \cite{Mattia}). Coefficients $c_{ls}$ and  $\gamma_1$ can be obtained from
 the substrate-liquid  surface free
energy expressed in the form \cite{deGennes1,gouin1}\,:
\begin{equation*}
\phi(\rho_{s})=-\gamma _{1}\rho_{s}+\frac{1}{2}\,\gamma_{2}\,\rho_{s}^{2}.
\label{surface energy}
\end{equation*}
Here $\rho_{s}\simeq \rho_l$ denotes the fluid density value at the surface; from the superficial tension $\sigma$ and Young's condition, we immediately get
 $\gamma_1 \approx 75$\,\  (\cite{gouin5}).
 \newline
In the upper graph of Fig.\,\ref{fig4} we present the free energy graph  $G(h_x)$.
Due to $h_x>({1}/{2})\,\sigma_l$, it is not  numerically possible to obtain  the limit
point  W  corresponding to the dry wall; point W is obtained by an
interpolation associated with the concave part of the  \emph{G}-curve. Point  P  follows from the drawing of the tangent line issued from  W   to the $G$-curve. In the lower graph of Fig.\,\ref{fig4}   we present the disjoining pressure graph $%
\Pi(h_x)$. The physical part of the disjoining pressure graph corresponding to $%
\partial\Pi/\partial h_x <0$  is associated
with  a liquid layer of several molecules thick. The part $%
\partial\Pi/\partial h_x >0$ is also  obtained by Derjaguin \emph{et al} \cite{Derjaguin}\,.
The reference length  $d$
is of the same order as $\sigma_l$   and is a good
length unit for very thin films. The total pancake thickness $e_{\rm p}$ is of
one nanometer order corresponding to a good thickness value for a
high-energy surface \cite{deGennes1,Israel}\,; consequently in the tall trees, at high level, the thickness of the layer is of a few
nanometers. The point P  of  the lower graph corresponds to the point P of upper graph. Let us note that the crude sap is not pure water; its liquid-vapour surface tension
has a lower value than the surface tension of pure water   and it is possible to obtain the same spreading coefficients
with less energetic surfaces. \newline
When $x_{_{\rm P}}$ corresponds to the altitude of the pancake layer,  $\Pi \simeq  \rho_l\,g\,x_{_{\rm P}}$  \cite{Derjaguin,gouin3}\,.  To this
altitude, we add 20 meters corresponding to the ascent of sap due to capillarity and osmotic pressure and we obtain on the lower graph of Fig.\,\ref{fig4} a maximum film height
of approximatively $140$ meters ($20+120$ meters) corresponding to $12$ atmospheres, which is of the same level order as the topmost trees.
\newline
These results arising from molecular physics require a comparison between the behaviours of liquid motions both in microtubes and in nanolayers :

When the xylem microtubes are  tight filled with crude sap, the liquid motions are   Poiseuille flows \cite{Zimm} and to be efficient the microtube radii should be as wide as possible, which is not the case. Due to the liquid
incompressibility  the flow is \emph{very rigid} and the pressure effects are fully propagated onto the microtube walls.

When the xylem microtubes are partially filled with thin water films, there are qualitative experiments for slippage on the walls when the film thickness  is of the mean free path  order \cite{Churaev,degennes}\,; the boundary condition  on the walls writes:
\begin{equation*}
\textbf{u}=L_{s}\, {d \textbf{u}}/{d n},
\end{equation*}
where  $\textbf{u}$ is the liquid velocity, ${d \textbf{u}}/{d n}$ is the normal derivative at the wall and $L_{s}$ is the so-called \emph{Navier length} \cite{degennes}\,. The Navier length   may be as large as a few microns \cite{Tabeling} and we obtained the  mean liquid velocity ${\mathbf{\overline{u}}}$  along a thin layer from   \cite{gouin5}\,:
\begin{equation}
\nu\, {\mathbf{\overline{u}}}=h_{x} \left( \frac{%
h_{x}}{3}+L_{s}\right) \left[ \,{\bf{grad}}\ \Pi (h_{x}) - g\,\textbf{i}\,
\right] ,  \label{variation potentiel chimique}
\end{equation}
where $\nu$ denotes the kinematic viscosity and $\textbf{i}$ the unit vertical vector.
Consequently  the slippage condition multiplies the flow rate by a factor of $(1+3L_{s}/h_{x})$. For example, if $h_{x}= 3\,nm$ and $%
L_{s}=100\,nm$, which is a Navier length of small magnitude with respect to
experiments, the multiplicative factor is $10^2$; if $L_{s}= 7\,\mu m$, as
considered in \cite{Tabeling}\,, the multiplicative factor is $10^{4}$ which is of the same order as for nanotube observations \cite{Mattia}\,.
\newline
Equation (\ref{variation potentiel chimique}) is   mainly realistic at the top of tallest trees where the xylem
network is strongly ramified; the heartwood  may
contain liquid under positive pressure and is connected with the sapwood \cite{Zimm}\,. The flow rate    can increase or decrease due to
the spatial derivative of the thickness $h_{x}$ and consequently depends on the local
disjoining pressure  value. The tree's  versatility adapts the disjoining
pressure gradient effects by opening or closing the stomatic pits, so that the bulk pressure in
micropores can be more or less negative and so, the transport of water is differently
dispatched in the stem parts.
 \\
The sap motion is induced by  the transpiration across
micropores located in tree leaves \cite{Zimm}\,. It seems natural to surmise
that the diameters of xylem  microtubes might result of a competition
between evaporation  which reduces the flow of sap  and the flux of
transpiration in micropores inducing the  motion. It is noticeable that if we replace the flat surfaces of the microtubes   with wedge
geometry  or  corrugated surface,  it is much easier to obtain the
complete wetting requirement;  thus,  plants can avoid having very high
energy surfaces. Nonetheless, they are still   internally wet  if crude sap flows  through
wedge shaped corrugated pores. The wedge does not have to be perfect on the
nanometric scale to significantly enhance the amount of liquid flowing at modest pressures, the walls    being
considered as  plane surfaces endowed with an average surface energy.
 \vspace{0.4cm}

\noindent \textbf{{Methods.}}
\\

\small
\noindent \emph{We compare two experiments:\\
- The Scholander pressure bomb experiment (1955) based on the cohesion-tension theory (1894) in which liquids are considered to be incompressible.\\
- The Sheludko experiment (1967) based on the concept of disjoining pressure in DLVO theory (1948) that highlights a strong difference between liquid bulk and thin layer pressures.\\
The theoretical results  allow us to obtain:\\
-  The computation of tallest trees' level that fits with real facts.\\
- The interpretation of the motion in xylem microtubes by using the shallow water approximation and the slippage on walls at the nanometric scale}.\\

\end{spacing}
\scriptsize
 \begin{spacing}{0.7}

 \end{spacing}

\begin{thebibliography}
\footnotesize{
\bibitem{Flindt}  {Flindt, R.} \emph{Amazing Numbers in
Biology} (Springer, New York, 2006).

\bibitem{Koch}  {Koch, W., Sillett, S.C., Jennings, G.M. \& Davis S.D.} The
limit to tree height. \emph{Nature}   \textbf{428},   851-854 (2004).

\bibitem{Zimm}  {Zimmermann, M.H. } \emph{Xylem Structure and the Ascent of
Sap}  (Springer, New York, 1983).


\bibitem{Dixon}  {Dixon, H.H. \& Joly, J.} On the ascent of
sap. \emph{Philosophical Transactions of the Royal Society of
London, B} \textbf{186}: 563-576 (1894).


\bibitem{Derjaguin}  {Derjaguin, B.V., Churaev, N.V. \& Muller, V.M. } \emph{Surfaces
Forces}  (Plenum Press, New York, 1987).

 \bibitem{gouin4} {Gouin, H.}   A mechanical model for the disjoining pressure, \emph{International Journal of Engineering Science}     \textbf{47},  691-699  (2009) \&  arXiv:0904.1809.



\bibitem{deGennes1}  {de Gennes, P.G.} Wetting : statics and
dynamics. \emph{Review of Modern Physics} \textbf{57},  827-863 (1985).



\bibitem{vanderHonert}  {van der Honert, T.H. } Water transport in plants as a
catenary process. \emph{Discussions of the Faraday Society}
\textbf{3}, 1105-1113 (1948).

\bibitem{Scholander}  {Scholander,
P.F., Love, W.E. \& Kanwisher, J.W. } The rise of sap in tall
grapevines. \emph{Plant Physiology} \textbf{30}, 93-104 (1955).


\bibitem{Haberlandt}  {Haberlandt, G. }
\emph{Physiological plant anatomy}  (MacMillan, London,  1914).

\bibitem{Foster}  {Foster, A.S. } Plant
idioblast: remarkable examples of cell specialization.
\emph{Protoplasma} \textbf{46}, 183-193 (1956).


\bibitem{Pridgeon} {Pridgeon, A.M.} Diagnostic
anatomical characters in the pleurothallidinae (orchidaceae).
\emph{American Journal of Botany} \textbf{69}, 921-938 (1982).


\bibitem{0'Brien} {0'Brien, T.P. \& Carr, D.J.
} A suberized layer in the cell walls of the bundle sheath of
grasses. \emph{Australian Journal of Biological Science}
\textbf{23}, 275-287 (1970).

\bibitem{Traube}   {Traube,  M. } Experimente zur
Theorie der Zellenbildung und Endosmose. \emph{Archiv f\"{u}r Anatomie, Physiologie und wissenschaftliche Medicin}, Berlin \textbf{34}, 87-165 (1867).



\bibitem{Preston}  {Preston, R.D. } In: Deformation and Flow in Biological
Systems
 (ed. Frey-Wyssling, A.)  257-321 , (North Holland Publishing, Holland, 1952).


\bibitem{Mackay}  {Mackay, J.F.G. \& Weatherley, P.E.} The effects of transverse cuts
through the stems of transpiring woody plants on water transport and
stress in the leaves. \emph{Journal of Experimental Botany}
\textbf{24}, 15-28 (1973).






\bibitem{Balling} {Balling, A.  \& Zimmermann, U.} Comparative
measurements of the xylem pressure of nicotiana plants by means of
the pressure bomb and pressure probe. \emph{Planta}  \textbf{182},
325-338 (1990).

\bibitem{Tyreeetal} {Tyree, M.T., Cochard, H. \& Cruiziat, P. } The
water-filled versus air-filled status of vessels cut open in air:
The 'Scholander assumption' revisited. \emph{Plant, Cell \&
Environment} \textbf{26}, 613-621 (2003).


\bibitem{Sheludko}  {Sheludko, A.} Thin liquid
films. \emph{Advance in Colloid Interface Science} \textbf{1},
391-464 (1967).






        \bibitem{gouin3}  {Gouin, H.} A new approach for the limit to tree
height using a liquid nanolayer model. \emph{Continuum Mechanics and
Thermodynamics} \textbf{20}, 317-329 (2008) \&  arXiv:0809.3529.


\bibitem{gouin5}  {Gouin, H.}  Solid-liquid interaction at nanoscale and its application in vegetal biology, \emph{Colloids and Surfaces A} \textbf{383}, 17-22 (2011) \&   arXiv:1106.1275.


\bibitem{rowlinson}   {Rowlinson J.S. \& Widom, B.} \emph{Molecular Theory of
Capillarity}  (Clarendon Press, Oxford, 1984).


\bibitem{gouin1}  {Gouin, H.} Energy of interaction between solid surfaces
and liquids. \emph{ The Journal of Physical Chemistry B}
\textbf{102},  1212-1218 (1998) \& arXiv:0801.4481.

\bibitem{Israel}   {Israelachvili, J.} \emph{Intermolecular Forces} (Academic Press, New York, 1992).

 \bibitem{Lifshitz}  {Dzyaloshinsky, I.E.,  Lifshitz, E.M. \&  Pitaevsky, L.P.}  The
general theory of van der Waals forces, \emph{Advances in Physics} \textbf{10}, 165-209 (1961).




\bibitem{Handbook} {Weast, R.C.}  (ed.)  \emph{Handbook of Chemistry and Physics}
65th Edition  (CRC Press, Boca Raton, 1984).


\bibitem{Mattia}  {Mattia, D. \& Gogotsi, Y.} Review: static and dynamic behavior
of liquids inside  carbon nanotubes, \emph{Microfluid Nanofluid} \textbf{5},  289-305 (2008).



\bibitem{Churaev}  {Chuarev, N.V.}  Thin liquid layers, \emph{Colloid J.} \textbf{58}, 681-693 (1996).

\bibitem{degennes} {de Gennes, P.G.}  On the fluid/wall slippage, \emph{Langmuir}  \textbf{8}, 3413-3414
(2002) \& arXiv:0115383.

\bibitem{Tabeling}  {Tabeling, P.}  (ed.)   Microfluidics, \emph{Comptes Rendus Physique},
\textbf{5},  2-608 (2004). }






\end{thebibliography}
 \end{document}